\documentclass{aa}
\usepackage{epsfig}

\begin{document}

\title{Internal shock model for Microquasars} 
\author{Christian R. Kaiser, Rashid Sunyaev and Henk C. Spruit}
\institute{Max-Planck-Institut f{\"u}r Astrophysik,
Karl-Schwarzschild-Str.1, 85740 Garching, GERMANY}
\offprints{ckaiser@mpa-garching.mpg.de}

\date{Received <date> / Accepted <date>}
\thesaurus{6(02.19.1; 08.02.3; 08.09.2 GRS 1915+105; 13.18.5)}

\maketitle
\begin{abstract}
We present a model for the radio outbursts of microquasars based on
the assumption of quasi-continuous jet ejection. The jets are `lit up'
by shock fronts traveling along the jets during outbursts. The shocks
accelerate relativistic particles which emit the observed synchrotron
radiation. The observed comparatively flat decay light curves combined
with gradually steepening spectral slopes are explained by a
superposition of the radiation of the aging relativistic particle
population left behind by the shocks. This scenario is the low energy,
time-resolved equivalent to the internal shock model for GRBs. We show
that this model predicts energy contents of the radiating plasma
similar to the plasmon model. At the same time, the jet model relaxes
the severe requirements on the central source in terms of the rate at
which this energy must be supplied to the jet. Observations of
`mini-bursts' with flat spectral slopes and of infrared emission far
from the source centre suggest two different states of jet ejections:
(i) A `mini-burst' mode with relatively stable jet production and weak
radio emission with flat spectra and (ii) an outburst mode with strong
variations in the jet bulk velocities coupled with strong radio
emission with steeper spectra. We also show that the continuous jets
in microquasars should terminate in strong shocks and possibly inflate
radio lobes similar to extragalactic jet sources. We investigate the
possibility of testing the predictions of this model with resolved
radio observations. Finally, we suggest that Doppler-shifted X-ray
iron lines, and possibly H-alpha lines, may be emitted by the jet flow
of microquasars if thermal instabilities analogous to those in SS433
exist in their jets.  
\end{abstract}

\section{Introduction}

The number of stellar mass black hole candidates known to produce jets
has increased considerably over recent years. While SS433 was for a
long time thought to be a rather exotic object, advances in X-ray
astronomy have dramatically increased the number of known galactic
X-ray binaries (van Paradijs 1995\nocite{jv95}). At least nine of
these sources subsequently showed evidence for the production of
relativistically moving jets. Mirabel \& Rodr{\' \i}guez
(1999)\nocite{mr99} give an excellent review on the observational and
theoretical status of these objects, the microquasars.

The presence of relativistically moving material in these sources was
discovered during radio observations. During times of strongly
enhanced radio emission, which in the following we will refer to as
radio outbursts, the emission region can often be resolved into at
least two components. These components are observed to separate in
opposite directions over a fews tens of days (i.e. Mirabel \&
Rodr{\'\i}guez 1994\nocite{mr94}). The projected velocity of the
component traveling on a trajectory towards the observer can exceed
the speed of light, if its intrinsic velocity is large (e.g. Rybicki
\& Lightman 1979\nocite{rl79}). In addition to the apparently
superluminal nature of the jet components, their propagation has been
observed to slow down only in one case (XTE J1748-288, Hjellming et
al. 1999\nocite{hrmshwgp99}). This constant expansion speed led to the
interpretation of practically ballistic trajectories of discrete
plasmon ejections as explanation for the observed radio components
(e.g. Mirabel \& Rodr{\'\i}guez 1999\nocite{mr99}).

Although suggested for the jets of SS433 (Hjellming \& Johnston
1988\nocite{hj88}), models with quasi-permanent jet production have
received little attention in the case of microquasars. This is
somewhat surprising given the large number of similarities they show
with jet producing extragalactic sources like quasars (hence the name
microquasars) and radio galaxies (e.g. Mirabel \& Rodr{\'\i}guez
1998\nocite{mr98}). In these cases there is little doubt that apart
from some possible minor intermittency of the jet production mechanism
for some sources (i.e. Reynolds \& Begelman 1997\nocite{rb97}) the jet
flow is practically continuous.

In this paper we endeavour to close this gap by the development of a
continuous jet model for microquasars to explain the radio outbursts
observed in these sources. The model is based on the idea of internal
shocks in jets (Rees 1978\nocite{mr78}) which was successfully applied
to Gamma Ray Bursts (GRB) (Rees \& Meszaros 1994\nocite{rm94}). In
Sect. \ref{sec:pj?} we briefly review the plasmon model and discuss
possible improvements on this within the jet picture. We develop the
treatment of the relativistic jet flow in Sect. \ref{sec:dyn} and
discuss the evolution of the synchrotron emission resulting from the
internal shock in Sect. \ref{sec:syn}. The model is then applied to
the probably best studied radio outburst of any microquasar, the March
1994 event in GRS 1915+105 (Mirabel \& Rodr{\'\i}guez
1994\nocite{mr94}) in Sect. \ref{sec:app}. The properties of the jet
of GRS 1915+105 like its energy content are derived from the model in
Sect. \ref{sec:prop}. In Sect. \ref{sec:end} we consider the
implications of a continuous jet for the interaction of microquasars
with their environment. Finally, in Sect. \ref{sec:test} we
summarise some observational consequences of our model which may be
used to test the model with future observations.

\section{Plasmons or jets?}
\label{sec:pj?}

Atoyan \& Aharonian (1999)\nocite{aa99} developed a model which is
intended to reconcile the idea of discrete magnetised plasmon
ejections during radio outbursts with the observed lightcurve and
spectral behaviour in the radio waveband. They find that a single
population of relativistic particles accelerated at the time of the
ejection of the plasmons cannot explain the observations. The energy
losses of the relativistic particles due to synchrotron radiation
would lead to a sharp cut-off in the radio spectrum moving to lower
frequencies as the plasmons expand and travel outwards. This is quite
different from the observed rather gentle steepening of the
spectrum. Atoyan \& Aharonian (1999)\nocite{aa99} also show that a
continuous replenishment of relativistic particles to the plasmons
alone cannot solve this problem because in this case the spectral
cut-off moves to higher frequencies with time. They therefore
postulate that the relativistic particles in the plasmons during the
March 1994 event in GRS 1915+105 were continuously replenished,
presumably by a shock at the side of the plasmons pointing towards the
source centre, but also suffered energy dependent escape losses.

To fit the observations the scenario proposed by Atoyan \& Aharonian
(1999)\nocite{aa99} requires that the mean free path of the most
energetic relativistic particles in the magnetised plasmons is
comparable to or exceeds the physical dimensions of the plasmons. This
implies that these particles travel through the plasmons producing
synchrotron emission but then leave them without scattering once off
irregularities in the magnetic field or other particles. This is
difficult to reconcile with the requirement that in order to be
accelerated to relativistic velocities in the shock regions the mean
free path in these regions must be short to ensure many shock
crossings. If the accelerating shocks are close to the plasmons, this
then means that the properties of the plasma change dramatically over
short distances. Moreover, it is not clear in this scenario why the
synchrotron emission is not completely dominated by the contribution
of the shocks themselves.

In this paper we propose a different scenario to explain the observed
properties of the radio emission of microquasars during
outbursts. This is based on the assumption that microquasars may
produce continuous jets for a long time before the actual outburst
occurs and may well do so permanently (see also Levinson \& Blandford
1996a, b)\nocite{lb96a}\nocite{lb96b}. The outbursts in our model
are then caused by two shocks traveling along these continuous jets
which accelerate the required relativistic particles in situ. After
the shock has passed a particular region in one of the jets, this
region continues to contribute to the total emission until the cut-off
in the specific spectrum of this region moves below the observing
frequency. The jet components observed in microquasars are in general
not resolved and so the measured flux is the integrated emission from
all the jet regions passed by the shock which are still emitting at
the relevant frequency. This implies that the observed spectrum is
steeper than that of the jet region immediately behind the shock where
radiation losses are still negligible. The variation of the strength
of the magnetic field and of the number of relativistic particles
accelerated by the shock along the jet then give rise to a slowly
steepening radio spectrum. This effect was discussed in the case of
the radio hot spots of powerful extragalactic radio sources by Heavens
\& Meisenheimer (1987)\nocite{hm87}.

Our investigation is based on the jet model of Blandford \& Rees
(1974)\nocite{br74} and Marscher \& Gear (1985)\nocite{mg85}. A
similar approach to explain the synchrotron self-Compton emission of
extragalactic jets was taken by Ghisellini et
al. (1985)\nocite{gmt85}. They develop a numerical scheme to follow
the evolution of the energy spectrum of the relativistic particles
downstream of the jet shock taking into account radiative as well as
adiabatic energy losses. They also include synchrotron self-absorption
in this calculation. Since we are mainly interested in the radio
emission of the jets of microquasars on rather large scales ($\sim
10^{14}$m), we can neglect any Compton scattering and absorption
effects on the energy spectrum. In this case only adiabatic and
synchrotron losses are important and we can use the analytic solution
for the evolution of the energy spectrum of the relativistic electrons
derived by Kaiser et al. (1997)\nocite{kda97a}.

The underlying physical processes of the model presented here are very
similar to the internal shock model proposed as explanation for GRB
(Rees \& Meszaros 1994\nocite{rm94}). The same scenario has also been
invoked to explain the X-ray and $\gamma$-ray emission of
extragalactic jets (Ghisellini 1999\nocite{gg99}). In the internal
shock model the energy of the shock traveling along the jet is thought
to be supplied by the collision of fast shells of jet material with
slower ones (Rees 1978\nocite{mr78}). In the case of GRB this energy
is released practically instantaneously leading to the extremely short
duration of the observed bursts of emission. In extragalactic jets the
shell collision may take longer but the large distance to these
objects makes it difficult to separate the contributions of multiple
collision to the total emission. As outlined above, we argue in this
paper that the jets of microquasars provide us with the possibility to
observe the development of internal shocks in jets resolved both in
space and time.

\section{Dynamics of the jet}
\label{sec:dyn}

We follow Marscher \& Gear (1985)\nocite{mg85} and Ghisellini et
al. (1985)\nocite{gmt85} in the assumption that all relevant physical
quantities of the jet and the jet material are simple power law
functions of the unprojected distance from the source centre, $R$. The
shock traveling along the jet will compress the jet material but we
will assume that it does not change the behaviour of the physical
quantities as a function of the distance from the source centre. The
radius of the cross section of the jet is assumed to follow $r \propto
R^{a_1}$. For a freely expanding, conical jet $a_1 =1$. The energy
density of the magnetic field as measured in the rest frame of the jet
material is then given by $u_{\rm B}'=u_{\rm B}' (R_{\rm o}) (
R/R_{\rm o} )^{a_2}$, where energy flux conservation would require
$a_2 =-2 a_1$. Here and in the following dashes denote quantities
measured in the rest frame of the jet material moving at relativistic
speeds while quantities measured in the frame of the observer are
undashed.

The two frames of reference, the rest frame of the observer and that
of the shocked jet material, are defined such that the origins of both
coincide when the radio outbursts starts, i.e. when the shocks are
formed in the centre of the source and start traveling
outwards. Consider a section of the jet after the shock has passed
through it. At time $t$ this section is located at

\begin{equation}
R_{\rm ob}=\sin \theta R = \sin \theta v_{\rm s} t / (1 \pm \beta
_{\rm j} \cos \theta),
\label{simple}
\end{equation} 

\noindent in the rest frame of the observer. Here, $\theta$ is the
angle of the jet to the line of sight and $R$ is the unprojected
distance of the jet section from the centre of the source, i.e. the
origin of the observer's rest frame. $v_{\rm s}$ is the deprojected velocity
of the shock as measured in the rest frame of the observer and $\beta
_{\rm j}=v_{\rm j} /c$ is the deprojected velocity of the shocked jet material in
this frame in units of the speed of light. The expression in brackets
in Eq. (\ref{simple}) takes account of the Doppler shifted time
measurements taken in the observer's frame caused by the receding
($+$) or approaching ($-$) component of the motion of the jet material
(e.g. Rybicki \& Lightman 1979\nocite{rl79}). Transforming $t$ and $R$
to the frame comoving with the jet material we find

\begin{eqnarray}
t' & = & \gamma _{\rm j} \left( \frac{t}{1\pm \beta _{\rm j} \cos
\theta} - \frac{v_{\rm j} R}{c^2} \right) \label{tdash}\\ 
R' & = &
\gamma _{\rm j} \left( R - \frac{v_{\rm j} t}{1\pm \beta _{\rm j} \cos
\theta} \right) 
\label{rdash},
\end{eqnarray}

\noindent where $\gamma _{\rm j}$ is the Lorentz factor corresponding
to the velocity of the shocked jet material, $v_{\rm j}$. For the
origin of the comoving rest frame, $R'=0$, we recover from Eqs.
(\ref{tdash}) and (\ref{rdash}) the well-known result
$t'=t\delta_{\pm}$ with the usual relativistic Doppler factor $\delta
_{\pm}= \left[ \gamma _{\rm j} \left( 1 \pm \beta _{\rm j} \cos \theta
\right) \right] ^{-1}$. Since the shock moves along the jet, most of
the observed emission is not produced at the origin of the comoving
frame but at $R'\ne 0$. Suppose the section of the jet introduced
above was passed by the shock at time $t'_{\rm s}$ as measured in the
frame comoving with the shocked gas. Since this section is at rest in
this frame, it is subsequently located at $R'_{\rm s}=v'_{\rm s}
t'_{\rm s}$, where $v'_{\rm s}$ is the velocity of the shock as
measured in the shocked gas' frame. Using this and substituting $R$
from Eq. (\ref{rdash}) in Eq. (\ref{tdash}) yields

\begin{equation}
t' = t \delta _{\pm} - \frac{v_{\rm j} v_{\rm s}' t_{\rm s}'}{c^2}.
\label{tcom}
\end{equation}

\noindent This expression illustrates the fact that the emission we
observe at a given time $t$ from different parts of the jet was not
produced simultaneously in the frame of the shocked gas.

We assume that only those parts of the jet contribute to the total
emission which have been passed by the shock. The emission regions
within the jet can therefore be labeled with their `shock time',
$t_{\rm s}'$, and are observed at the intrinsic time $t'$ given by
Eq. (\ref{tcom}). Since $t' \ge t_{\rm s}' \ge 0$, this implies
$t \delta _{\pm} c^2/ (c^2+v_{\rm j} v_{\rm s}') \le t' \le t \delta
_{\pm}$. For convenience we introduce the ratio $\tau (t') = R/R_{\rm
o} = \gamma _{\rm j} ( v_{\rm s}' t_{\rm s}' + v_{\rm j} t') / R_{\rm
o}$, where we have used Eqs. (\ref{tdash}) and (\ref{rdash}).

\section{Synchrotron emission of the jet}
\label{sec:syn}

\subsection{Energy losses of the relativistic electrons}

We assume that the shock passing through the jet material accelerates
a population of relativistic electrons and/or positrons. During the
acceleration process and afterwards these relativistic particles are
subject to energy losses due to the approximately adiabatic expansion
of the jet material and synchrotron radiation. To determine the exact
form of the energy spectrum of the relativistic particles in a given
jet region the kinetic equation including acceleration and energy
terms must be solved. Heavens \& Meisenheimer (1987)\nocite{hm87}
present analytic and numerical solutions for some simplified
cases. They find that the energy spectrum follows a power law with a
high energy cut-off. The cut-off occurs at the energy for which energy
gains due to shock acceleration balance the synchrotron energy losses
(e.g. Drury 1983\nocite{ld83}). The cut-off becomes steeper further
downstream from the shock. For simplicity we assume that the
relativistic particles in a given jet region are initially accelerated
at a time $t'_{\rm s}$ during a short time interval $dt'_{\rm s}$ to a
power law spectrum with a sharp high energy cut-off at $\gamma _{\rm
max} (t'_{\rm s})$. In terms of the number of particles this can be
expressed by

\begin{eqnarray}
N' (\gamma _{\rm s}) d\gamma _{\rm s} & = & \left\{
\begin{array}{ll}
\dot{N}'_{\rm o} (t'_{\rm s}) \gamma _{\rm s} ^{-p} d\gamma _{\rm s}
dt'_{\rm s} & ; \gamma_{\rm s} \le
\gamma _{\rm max}\\
0 & ; \gamma _{\rm s} > \gamma _{\rm max}.
\end{array}
\right.
\label{spec}
\end{eqnarray}

\noindent Here $N'_{\rm o}(t'_{\rm s})$ is the rate at which
relativistic particles are accelerated in the jet by the shock at time
$t'_{\rm s}$. The normalisation of this energy spectrum and the
position of the high energy cut-off depend on the local conditions for
diffusion in the jet (e.g. Drury 1983\nocite{ld83}). These are not
straightforward to estimate and we therefore assume for simplicity
that the initial high energy cut-off of the relativistic particles
freshly accelerated at time $t'_{\rm s}$ is independent of $t'_{\rm
s}$. In the internal shock model for GRB the shock is caused by the
collision of shells of jet material moving at different
velocities. For GRB it is implicitly assumed that the collision energy
is dissipated very close to instantaneously. In the case of
microquasars the propagation of the shock is resolved in time. For the
normalisation of Eq. (\ref{spec}) we therefore assume

\begin{equation}
\dot{N}'_{\rm o} (t'_{\rm s})= \dot{N}'_{\rm o} (R_{\rm o}) e^{-R_{\rm
s}/(R_{\rm o} a_4)},
\label{expo}
\end{equation}

\noindent where $R_{\rm s}$ is the position of the shock at time
$t'_{\rm s}$ and $a_4$ a model parameter. This implies that the rate
at which the collisional energy is dissipated and partly conferred to
the relativistic particles is almost constant for $R_{\rm s}/R_{\rm o}
< a_4$ and decreases exponentially at larger distances. The onset of
the exponential behaviour then signifies the point at which almost all
of the collisional energy has been dissipated and the shock starts to
weaken significantly. This would coincide with the time at which the
two colliding shells have practically merged into one. Alternatively,
in only intermittently active sources the exponential decrease may be
caused by the shock, and therefore the fast shell causing the shock,
reaching the end of the jet.

After the passage of the shock the relativistic particles continue to
loose energy. The rate of change of the Lorentz factor of these
particles due to the nearly adiabatic expansion of the jet is given by
(e.g. Longair 1981\nocite{ml81})

\begin{equation}
\frac{d\gamma}{dt'} = -\frac{\gamma}{3 \Delta V'} \, \frac{d \Delta V'}{dt'},
\end{equation}

\noindent where $\Delta V'$ is the volume of the jet region the
particles are located in. We assume that the bulk velocity of the
shocked jet material is constant and this implies that the jet is only
expanding perpendicular to the jet axis, i.e. $\Delta V' \propto
\left( R/R_{\rm o} \right)^{2a_1} = \tau ^{2 a_1}$. Changing variables
from $t'$ to $\tau'$ then yields

\begin{equation}
\frac{d\gamma}{d\tau'} = - \frac{2 a_1}{3} \, \frac{\gamma}{\tau'}.
\label{adlo}
\end{equation}

\noindent Energy losses due to synchrotron radiation give

\begin{equation}
\frac{d\gamma}{d\tau'} = - \frac{4}{3} \, \frac{\sigma _{\rm
T}}{m_{\rm e} c^2} \, \gamma ^2 u_{\rm B}' (R_{\rm o}) \, \frac{R_{\rm
o}}{\gamma _{\rm j} v_{\rm j}} \tau'^{-2 a_1},
\label{sylo}
\end{equation}

\noindent where $\sigma_{\rm T}$ is the Thompson cross section and
$m_{\rm e}$ the rest mass of an electron. By summing Eqs.
(\ref{adlo}) and (\ref{sylo}) and integrating we find the Lorentz
factor $\gamma$ at time $t'$ of those electrons which had a Lorentz
factor $\gamma _{\rm s}$ at time $t'_{\rm s}$ (see also Kaiser et
al. 1997\nocite{kda97a})

\begin{equation}
\gamma (t', t'_{\rm s}) = \frac{\gamma _{\rm s} \tau'(t')^{-2/3
a_1}}{\tau'(t'_{\rm s}) ^{-2/3 a_1} +b_1 (t',t'_{\rm s}) \gamma _{\rm s}
},
\label{evol}
\end {equation}

\noindent with

\begin{equation}
b_1(t',t'_{\rm s})= \frac{4}{3 a_3} \, \frac{\sigma _{\rm T}}{m_{\rm e}
c} \, \frac{u_{\rm B}(R_{\rm o}) R_{\rm o}}{\gamma _{\rm j} v_{\rm j}}
\left[ \tau'(t')^{a_3} - \tau'(t'_{\rm s})^{a_3} \right]
\end {equation}

\noindent and

\begin{equation}
a_3 = 1- 2 \left( a_2 + \frac{a_1}{3} \right).
\end{equation}

\noindent The number of relativistic particles with a Lorentz factor
in the range $\gamma$ to $\gamma + d\gamma$ in the jet region
overtaken by the jet shock at time $t'_{\rm s}$ is therefore given by

\begin{eqnarray}
N'(\gamma) d\gamma & = & \left\{
\begin{array}{ll}
\dot{N}_{\rm o}' (t'_{\rm s}) \gamma ^{-p} b_2 & \\ \times
\tau'(t')^{-2/3 a_1} d\gamma dt'_{\rm s} & ; \gamma \le \gamma _{\rm
max} (t')\\ 0 & ; \gamma > \gamma _{\rm max} (t'),
\end{array}
\right.
\label{cspec}
\end{eqnarray}

\noindent with

\begin{equation}
b_2=\left[ \tau'(t')^{-2/3 a_1} - b_1 (t',t'_{\rm s}) \gamma
\right]^{p-2} \tau'(t'_{\rm s})^{2/3 a_1 (p-1)}
.
\end{equation}

\noindent Note here that the high energy cut-off, $\gamma _{\rm max} (t')$, also evolves according to Eq. (\ref{evol}).

\subsection{Synchrotron emission}

The synchrotron emission of the relativistic particles at time $t'$ in
the region of the jet overtaken by the shock at time $t'_{\rm s}$ is
given by

\begin{equation}
dP'_{\nu'} = \int _{\gamma _{\rm min}} ^{\gamma _{\rm max}}
\frac{4}{3} \sigma _{\rm T} c u_{\rm B} (t',t_{\rm s}') \gamma ^2 \Phi (\nu', \gamma) N'(\gamma) d\gamma,
\label{syn}
\end{equation}

\noindent where $\Phi (\nu', \gamma)$ is the synchrotron emission
spectrum of a single electron with Lorentz factor $\gamma$. Here we
assume that the magnetic field in the jet is tangled on scales smaller
than the radius of the jet. This then implies that $\Phi (\nu',
\gamma) = \bar{F} (\gamma)/ \nu _{\rm c}$, where $\bar{F}$ is one of
the synchrotron integrals normalised to give $\int _1 ^{\infty}
\bar{F} (\gamma) d\gamma = 1$ (e.g. Shu 1991\nocite{fs91}). To get the
total emission of the jet behind the jet shock we have to sum the
contributions of all the regions labeled with their shock times
$t'_{\rm s}$ within the jet. From Eq. (\ref{cspec}) we see that
this implies integrating Eq. (\ref{syn}) over $t'_{\rm s}$. This
integration must be performed numerically. Finally, to compare the
model results with the observations we have to transform to the rest
frame of the observer, $P_{\nu} = P'_{\nu'} \delta_{\pm}^3$. Note here
that the luminosity inferred from the observations at frequency $\nu$
was emitted in the gas rest frame at a frequency $\nu'=\nu
/\delta_{\pm}$.

As mentioned above, the steepening of the radio spectrum of the
superluminal jet components observed during outbursts of microquasars
is caused by the integrated emission from an extended region of the
jet. The further away from the shock the emission is created in the
jet, the lower the cut-off in the energy spectrum of the relativistic
electrons will be. However, if the jet region contributing to the
total emission is not too large the properties of the energy spectrum
will not change dramatically within this region. In this case we may
estimate the approximate location of the break in the radio spectrum
beyond which the spectrum steepens significantly. Most of the energy
lost by relativistic particles is radiated at their critical
frequency, $\nu _{\rm c} = \left( 3/2 \right) \nu _{\rm L} \gamma ^2$,
where $\nu _{\rm L}$ is the Larmor frequency. Defining the break
frequency, $\nu_{\rm b}$, as the critical frequency of the most
energetic particles just behind the jet shock we get

\begin{equation}
\nu' _{\rm b} = \frac{3 q \sqrt{ 2 \mu _{\rm o} u'_{\rm B}}}{4 \pi
m_{\rm e}} \gamma _{\rm max}^2,
\label{break}
\end{equation}

\noindent where $q$ is the elementary charge and $\mu _{\rm o}$ is the
magnetic permeability of the vacuum. The steepening of the radio
spectrum of the observed outbursts of microquasars strongly suggest
that the observing frequency, $\nu$, is close to the break frequency,
i.e. $\nu \sim \nu'_{\rm b} \delta _{\pm}$. Therefore, if most of the
observed emission comes from the region just behind the shock, we
expect from Eq. (\ref{break}) that $\gamma _{\rm max} \propto
u'_{\rm B} (R_{\rm o})^{-1/4}$. This implies that $\gamma _{\rm max}$
and $u'_{\rm B}(R_{\rm o})$ are not independent parameters of the
model but that they are correlated.

\section{Application to GRS 1915+105}
\label{sec:app}

The number of parameters in the model outlined above is large. In
order to reduce this number we assume that the jets in microquasars
are freely expanding, i.e. $a_1=1$, and that the flux of magnetic
energy through the jet is conserved, i.e. $a_2=-2$. Since we assume
the bulk velocity of the jet material to be constant, this implies
that the ratio of the kinetic energy and the energy of the magnetic
field is constant as well. Furthermore, we impose symmetry between the
approaching and receding sides of the source in the sense that the
model parameters describing the jet are the same on both sides. This
may be a poor assumption as the jets of GRO J1655-40 are observed to
be asymmetric (Hjellming \& Rupen 1995\nocite{hr95}). The model then
depends on five free parameters: The e-folding distance of the number
of relativistic particles within the jet accelerated by the shock,
$a_4$, the bulk velocity of the shocked jet material, $v_{\rm j}$, the
maximum Lorentz factor up to which relativistic particles are
initially accelerated, $\gamma _{\rm max}$, the slope of the initial
power law energy spectrum of these particles, $p$, and the energy
density of the magnetic field at $R_{\rm o}$, $u'_{\rm B} (R_{\rm
o})$. The acceleration rate of relativistic particles at the
normalisation radius, $\dot{N}'_{\rm o}(R_{\rm o})$, is in principle
also a free parameter. However, from Eq. \ref{syn}) we note that
it is only a multiplicative factor in the calculation of the total
radio emission of the jet. We therefore use it to normalise the model
in such a way that for a given set of model parameters $\dot{N}'_{\rm
o}(R_{\rm o})$ is such that the difference between the model
predictions and the observational data is smallest for this given set
of parameters.

Many radio outbursts of a number of microquasars have been
observed. But to constrain the model parameters in a meaningful way,
we would ideally need radio observations at two or more frequencies
which clearly resolve the approaching and the receding jet
component. Furthermore, the resolution should be sufficient to decide
whether one of these jet components consists of multiple
subcomponents, i.e. multiple shocks, the emission of which may be
blended in observations of lower resolution. To date there are very
few simultaneous multi-frequency observations which come even close to
this ideal situation. The best studied radio outburst of any
microquasar is still that of March 19th 1994 of GRS 1915+105 (Mirabel
\& Rodr{\'\i}guez 1994\nocite{mr94}). This is also the outburst studied
by Atoyan \& Aharonian (1999)\nocite{aa99}. To test our model we will
use the comparatively large data base accumulated during this event.

\begin{figure}
\centerline{
\epsfig{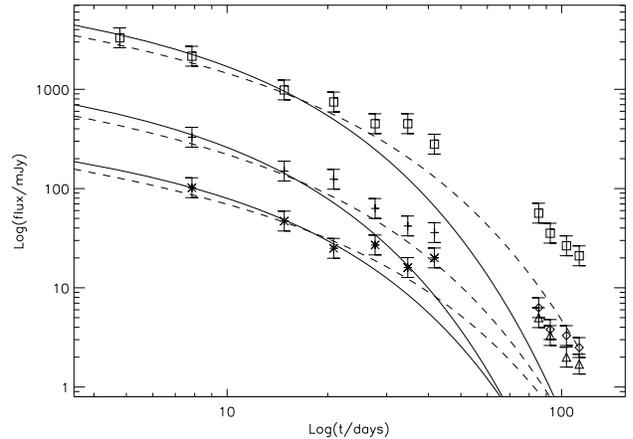}}
\caption{Flux density measurements and model fits of the March 19
radio outburst of GRS 1915+105 at 8.4 GHz. Crosses and diamonds:
Approaching jet component during first and second VLA campaign
respectively, stars and triangles: receding jet component, same
observations. Squares: Total radio flux at 8.4 GHz multiplied by a
factor 5 for clarity. Solid lines: Predictions of the fiducial model
fitted excluding flux measurements at 3.3 GHz for (from top to bottom)
the total flux, the approaching and the receding jet component. Dashed
lines: like solid lines but using all data for the model
fit. Observations from Mirabel \& Rodr{\'\i}guez (1994) and
Rodr{\'\i}guez \& Mirabel (1999).}
\label{fig:res8.4}
\end{figure}

\subsection{The observations}

The radio outburst of GRS 1915+105 which occurred in March 1994 was
one of the strongest recorded for this object. The flux density at 1.4
GHz exceeded 1 Jy which is at least ten times higher than the radio
flux in quiescence (Rodr{\'\i}guez et al. 1995\nocite{rgmgv95}). The
outburst was observed with the VLA in A-array at 8.4 GHz during 7
epochs covering almost 42 days. Except for the first of these, the two
jet components were resolved at this frequency (Mirabel \&
Rodr{\'\i}guez 1994\nocite{mr94}). For the first unresolved and one
further epoch measurements with the VLA are also available at 4.9 GHz
and 15 GHz. In addition, GRS 1915+105 was monitored during this time
by the Nancay telescope at 1.4 GHz and 3.3 GHz (Rodr{\'\i}guez et
al. 1995\nocite{rgmgv95}). There are 24 flux measurements at each
frequency but the source is unresolved at these frequencies. After a
reconfiguration of the VLA four more observations of GRS 1915+105 of
lower resolution were obtained at 8.4 GHz in B-array (Rodr{\'\i}guez
\& Mirabel 1999\nocite{rm99}). These measurements cover roughly
another 28 days but there is a gap of about 40 days between the end of
the A-array observations and the start of the B-array campaign.

From the resolved VLA observations at 8.4 GHz Mirabel \&
Rodr{\'\i}guez (1994)\nocite{mr94} determined a velocity for the jet
components of 0.92 c and an angle of the jets to the line of sight of
70$^{\circ}$. This assumes that the approaching and the receding
component travel at the same velocity in opposite
directions. Furthermore, it is assumed that GRS 1915+105 is located
12.5 kpc away from us. Fender et al. (1999)\nocite{fgmmpssw99}
observed another radio outburst of GRS 1915+105 in October 1997 and
found a higher intrinsic velocity which is inconsistent with a
distance of 12.5 kpc. They argue that the most likely distance for
this object is 11 kpc which then implies that the velocity of the jet
components in March 1994 was 0.86 c and the angle of the jets to the
line of sight is 68$^{\circ}$. In the following we will adopt these
later values. Extrapolating back the trajectories of the two jet
components Mirabel \& Rodr{\'\i}guez (1994)\nocite{mr94} find that the
outburst started at 20 hours on March 19.

\begin{figure}
\centerline{
\epsfig{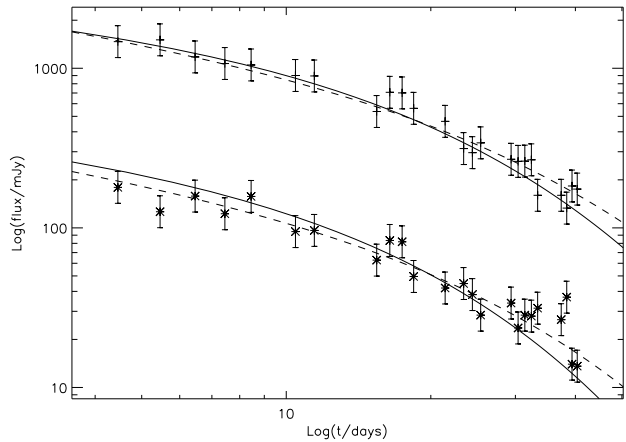}}
\caption{Flux density measurements of the March 19 radio outburst of
GRS 1915+105 at 1.4 GHz and 3.3 GHz. Crosses: Total radio flux at 1.4
GHz, stars: Total radio flux at 3.3 GHz divided by a factor 5 for
clarity. Model predictions as in Fig.
\ref{fig:res8.4}. Observational data from Rodr{\'\i}guez et
al. (1995).}
\label{fig:res1.4}
\end{figure}

Figs. \ref{fig:res8.4} and \ref{fig:res1.4} show all available flux
density measurements as a function of time. Rodr{\'\i}guez \& Mirabel
(1999)\nocite{rm99} note that another outburst of GRS 1915+105
occurred on April 21. The jet components of this new outburst are
clearly visible as an unresolved emission peak coincident with the
source centre in the VLA radio map of epoch 6. The approaching
component of this new outburst is also distinctly visible on the map
of the following observing epoch while the receding component is
probably blended with that of the previous outburst of March 19. The
signature of this later outburst as a sudden increase of the radio
flux is not very distinct at 1.4 GHz and 3.3 GHz. Even at 8.4 GHz the
situation in terms of the total flux density is somewhat
unclear. However, the outburst can be easily identified as a separate
event from the one of March 19 because the VLA maps reveal an emission
peak distinct from those of the earlier outburst. The later
observation epochs at the VLA with lower resolution detect the jet
components of the April 21 burst while the components of the March 19
event were not detected (Rodr{\'\i}guez \& Mirabel
1999\nocite{rm99}). This is rather puzzling as the extrapolation of
their lightcurves from the earlier observations indicate that they
should still have been visible during these later observations.

\subsubsection{The blending of outbursts}

A close examination of the total radio flux at 8.4 GHz reveals another
sudden increase around observation epoch 4 (April 9, Fig.
\ref{fig:res8.4}). This increase is also detected at 1.4 GHz and 3.3
GHz by the Nancay telescope on April 5 (Rodr{\'\i}guez et
al. 1995\nocite{rgmgv95}, Fig. \ref{fig:res1.4}). Atoyan \&
Aharonian (1999)\nocite{aa99} point out that this may be caused by a
sudden additional injection of fresh relativistic electrons in the
`blobs' of gas they consider in their model. An alternative
interpretation, which we will adopt here is that another, smaller
outburst occurred shortly after April 4. In this case, the radio
emission caused by the propagation of a new shock down the jet is most
likely blended with that of the earlier event of March 19 because of
the limited resolution of the observations. If the smaller outburst
occurred at April 5, 0:0 hours, and assuming that the apparent shock
velocity of this outburst is equal to that of the March 19 event,
i.e. 17.5 mas day$^{-1}$ (Rodr{\'\i}guez \& Mirabel
1999\nocite{rm99}), then the distance of the shock from the source
centre on the approaching jet side at observing epoch 4 would be
roughly 0.1". The position of the emission peak on April 9 is given as
0.36" from the source centre (Mirabel \& Rodr{\'\i}guez
1994\nocite{mr94}). The resolution of the VLA in A-array at 8.4 GHz is
$\sim$0.3" and this means that a secondary peak caused by a later,
somewhat weaker outburst as proposed here could not be detected as an
individual structure. The situation on the receding jet side with its
lower expansion velocities is even worse. However, the radio emission
of the additional outburst would contribute to the total radio flux of
the source and we believe that this has been detected here.

In summary, we will assume that the observations outlined above cover
three separate radio outbursts of GRS 1915+105. The first and
strongest occurred on March 19. The jet components of this burst are
detected until the end of the first set of VLA observations (April
30). By the time of the second VLA campaign (June 13) they had
vanished although the extrapolation of their earlier lightcurves
suggested that they should still be observable. The second much weaker
outburst occurred shortly after April 4, since the Nancay data show a
sudden increase in radio flux on April 5 but not on April 4. The radio
emission of this event is most likely blended with that of the first
outburst and there is no sign of this second burst in the second set
of VLA observations. The third outburst finally started on April 21
and was intermediate in strength. The radio emission caused by this
event is clearly detected in the source centre on April 23 and the
approaching jet component can be seen in the VLA map obtained on April
30. Both jet components are clearly detected in all four VLA observing
epochs in June and July.

\subsubsection{Data used in the modeling}

In the continuous jet model for microquasars outlined above, only one
shock is thought to travel outwards in each jet. Because of this, only
observational data from observing epochs during which we can be sure
that there was only one shock per jet contributing to the radio
emission can be used in constraining the model parameters. From the
above discussion it is clear that we can only use the three unresolved
VLA measurements at 4.9 GHz, 8.4 GHz and 14.9 GHz from March 24 and
the two following resolved observations at 8.4 GHz. These later
observation provide us with separate flux measurements for the
approaching and receding jet components. We also use the measurement
of the receding jet component of April 9, since it seems likely that
all the flux of the second outburst was attributed to the approaching
component by Mirabel \& Rodr{\'\i}guez
(1994)\nocite{mr94}. Alternatively, this measurement can be taken as
an upper limit. Of the Nancay observations we use all flux
measurements starting March 24 through to April 4. There are eight
observations during this time at 1.4 GHz and 3.3 GHz. For all 24
measurements used to constrain the model we assume the conservative
error of 46\% suggested by Rodriguez \& Mirabel (1999)\nocite{rm99} as
opposed to the original error of 5\% quoted by Rodr{\'\i}guez et
al. (1995)\nocite{rgmgv95}. In practice we did not use the eight 3.3
GHz data points as their inclusion led to significantly worse fits of
the model to the observational data. See the next section for a
discussion of this point.

\subsection{Constraining the model parameters}

We use the model outlined above to calculate the expected radio flux
at the times GRS 1915+105 was observed during the outburst starting
March 19. The `goodness of fit' of the model for a given set of free
parameters to the observations was assessed by calculating the sum of
the $\chi^2$-differences at the times the source was observed. The
best-fitting model was then found by minimising this $\chi ^2$-value
using a 4-dimensional downhill simplex method (Press et
al. 1992\nocite{ptvf92}). The remaining fifth model parameter, the
energy density of the magnetic field at the normalisation radius,
$u'_{\rm B}(R_{\rm o})$, was set `by hand' to five different
values. The normalisation radius, $R_{\rm o}$, was set to $1.1\cdot
10^{14}$ m which corresponds to the unprojected distance of the jet
shock from the source centre at the time of the first VLA observation,
i.e. March 24.

The results of the model fits are summarised in Table
\ref{tab:res}. The model fits the observational data equally well for
all five adopted values for the strength of the magnetic field. As
expected from Eq. (\ref{break}) we find that the maximum Lorentz
factor up to which relativistic particles are accelerated correlates
strongly with the value of the energy density of the magnetic
field. The values found for $\gamma _{\rm max}$ in the model fits
follow almost exactly a $u'_{\rm B}(R_{\rm o})^{-0.25}$ law. This
shows that the model presented here cannot be used to constrain the
strength of the magnetic field within the jet independently of the
maximum energy of the relativistic particles. To proceed we adopt in
the following $u'_{\rm B}(R_{\rm o})=4.2\cdot 10^{-5}$ as our fiducial
model. This corresponds to the equipartition value of the magnetic
field of $10^{-5}$ T (0.1 G) found by Atoyan \& Aharonian
(1999)\nocite{aa99} for the chosen value of $R_{\rm o}$. Note however,
that the energy density of the magnetic field and that of the
relativistic particles does not stay in equipartition for all times in
our model.

\begin{table*}
\begin{center}
\caption{Results of the model fits. The first five fits are obtained
without the 3.3 GHz data points. $u'_{\rm B}$ is the energy density of
the magnetic field, $a_4$ is the radius beyond which the acceleration
rate of relativistic particles decreases exponentially in units of
$R_{\rm o}$ (see Eq. \ref{expo}), $v_{\rm j}$ is the bulk
velocity of the shocked jet material, $\gamma _{\rm max}$ is the
initial high energy cut-off of the energy spectrum of the relativistic
particles, $p$ is the initial slope of this spectrum and
$\dot{N}'_{\rm o}$ is its normalisation.}
\label{tab:res}
\begin{tabular}{cccccccc}
& $u'_{\rm B}(R_{\rm o}) / \mbox{J m$^{-3}$}$ & $a_4$ & $v_{\rm j}$ / c & $\log(\gamma
_{\rm max})$ & $p$ & $\chi ^2$ & $\dot{N}'_{\rm o}(R_{\rm o})$\\
\hline\\
& $4.2 \cdot 10^{-7}$ & 2.8 & 0.61 & 3.2 & 1.4 & 0.26 & $2.3\cdot
10^{41}$\\
& $1.1 \cdot 10^{-5}$ & 2.7 & 0.61 & 2.9 & 1.4 & 0.26 & $5.5 \cdot
10^{40}$\\
without 3.3 GHz data & $4.2\cdot 10^{-5}$ & $2.7^{+4.9}_{-1.0}$ & $0.61^{+0.14}_{-0.13}$ & $2.7^{+0.8}_{-0.36}$ & $1.4^{+0.8}_{-0.9}$ &
0.26 & $1.5 \cdot 10^{40}$\\
& $1.1 \cdot 10^{-3}$ & 2.7 & 0.61 & 2.4 & 1.4 & 0.26 & $2.1 \cdot
10^{39}$\\
& $4.2 \cdot 10^{-3}$ & 2.6 & 0.61 & 2.2 & 1.4 & 0.26 & $9.5 \cdot
10^{38}$\\[1.5ex]
all data & $4.2\cdot 10^{-5}$ & 4.2 & 0.56 & 2.8 & 1.9 & 1.02 & $1.2
\cdot 10^{41}$\\[0.5ex]
\hline
\end{tabular}
\end{center}
\end{table*}

Figs. \ref{fig:res8.4} and \ref{fig:res1.4} show the model
predictions of the fiducial model compared to the observational
data. Also shown are the predictions of the model for the best-fitting
parameters when using also the 3.3 GHz data. If we use all available
data, the model predictions at the higher observing frequencies are
rather low at early times. For the VLA measurements at 4.9 GHz and
14.9 GHz (not shown in Figs. \ref{fig:res8.4} and \ref{fig:res1.4})
on March 24 the model predicts flux densities of 759 mJy and 385 mJy
respectively. This is much lower than the measured flux densities of
887 mJy and 514 mJy at these frequencies. A closer inspection of
Fig. \ref{fig:res1.4} also shows that the measurement at 4.9 GHz
actually exceeds those taken at 3.3 GHz at comparable times, which is
unlikely to be real. Table \ref{tab:res} also shows that the fit
obtained including the 3.3 GHz data is much worse in terms of the
reduced $\chi ^2$-values than that excluding them. Also the flux
densities predicted by our fiducial model which excludes the 3.3 GHz
data, 926 mJy at 4.9 Ghz and 499 mJy at 14.9 Ghz, are much closer to
the observations. Finally, the predictions of this model at 3.3 GHz
fit the observations well at this frequency apart from the early
observing epochs (see Fig. \ref{fig:res1.4}). We therefore believe
that the model parameters found using our fiducial model and excluding
the 3.3 GHz data are more reliable than those found when including
these additional measurements.

To estimate the expected error of the model parameters of our fiducial
model we calculated the $\chi ^2$-value for a large set of
combinations of the 4 free model parameters. The uncertainties quoted
in Table \ref{tab:res} are 1-$\sigma$ errors corresponding to those
parameter ranges for which $\chi ^2 \le 1$. Note that the
uncertainties of the model parameters, particularly those of $a_4$ and
$p$, are large while the light curves predicted by the model pass the
data points well within the error bars of the flux measurements (see
Figs. \ref{fig:res8.4} and \ref{fig:res1.4}). This suggests that the
quoted errors of the observed fluxes, at least for the VLA data
points, are too conservative which also results in an overestimation
of the uncertainties of the model parameters.

\subsection{Comparison with the later observations}

The exponential function which describes the change in the
acceleration rate of relativistic particles as a function of the
position of the shock in the jet, Eq. (\ref{expo}), implies that the
radio flux caused by the first outburst on March 19 decreases quickly
once $R_{\rm s} / R_{\rm o}$ exceeds $a_4$. This effect is clearly
visible in Fig. \ref{fig:res8.4} at 8.4 GHz. The model predicts that
without the additional blended radio emission caused by the second and
third outbursts around April 5 and April 21 the two jet components
would have faded much more rapidly than is observed. This effect is
less pronounced at lower frequencies (see
Fig. \ref{fig:res1.4}). However, even for these the predicted and the
observed light curves steepen somewhat roughly 15 days after the start
of the first outburst. The continued steepening of the lightcurves of
the two jet components at 8.4 GHz can also explain why they were not
detected during the second observing campaign at the VLA (see Fig.
\ref{fig:res8.4}).

We tried replacing the exponential in Eq. (\ref{expo}) with a
simple power law. The data can be adequately fitted with this modified
model as well. However, this change in the temporal behaviour of
$\dot{N}'_{\rm o}(R_{\rm o})$ also leads to a much increased flux at
low observing frequencies at later times. Using this modified model we
found a flux at 1.4 GHz 40 days after the start of the first outburst
exceeding the observations by a factor of at least 1.3 even without
considering the possible contribution from later outbursts. This
supports the picture of two colliding shells of jet material of finite
width causing the internal shock. The rate at which energy is
dissipated is roughly constant during the collision and decreases
rapidly once the two shells have merged.

Comparing the model lightcurves with the observational data the
signature of the second outburst starting around April 5, about 16
days after the start of the first outburst, can clearly be detected at
8.4 GHz. The increase in the radio emission caused by the second event
is less dramatic at 3.3 GHz and 1.4 GHz but can still be seen in
Fig. \ref{fig:res1.4}. The third outburst of April 21, 32 days after
the first burst, is seen as excess emission at 8.4 GHz and 3.3 GHz but
is less obvious at 1.4 GHz. We note that in general the smooth
lightcurve predicted by our model fits the VLA observations much
better than the flux measurements at 1.4 GHz and 3.3 GHz taken with
the Nancay telescope. This may imply larger errors for the low
frequency data which hide to some extent the signatures of the second
and third outburst which are much weaker than the first.

\section{Properties of GRS 1915+105}
\label{sec:prop}

\subsection{Energetics}

Using the model parameters of our fiducial model we now derive some of
the physical properties of the jets of GRS 1915+105 during the
outburst of March 19. The rate at which energy is transfered by the
jet shock to the relativistic particle population at time $t'_{\rm s}$
is

\begin{eqnarray}
\dot{E}'_{\rm rel} & = & \dot{N}'_{\rm o}(R_{\rm o}) e^{-R_{\rm s}/(R_{\rm
o} a_4)} m_{\rm e} c^2 \int _{\gamma _{\rm min}}^{\gamma _{\rm max}}
\gamma _{\rm s}^{-p} (\gamma _{\rm s} -1) \, d\gamma _{\rm s}
\nonumber\\
& = & 10^{29} e^{-R_{\rm s}/(R_{\rm o} a_4)} \mbox{W},
\end{eqnarray}

\noindent where we have assumed that only electrons and/or positrons
are accelerated and that the initial energy spectrum of the
relativistic particles extends down to $\gamma _{\rm min} =1$. 

The rate at which energy is transported in the form of magnetic fields
can be estimated by

\begin{eqnarray}
\dot{E}'_{\rm B} & \approx & \pi \left( \frac{\theta R_{\rm s}}{2}
\right)^2 u'_{\rm B}(R_{\rm s}) v'_{\rm s} \nonumber\\
& = & 1.8 \cdot 10^{28} \left(
\frac{\theta}{\mbox{degrees}} \right) ^2 \mbox{W},
\end{eqnarray}

\noindent where $\theta$ is the opening angle of the conical jet and
$v'_{\rm s}$ is the speed of the shock in the frame of the shocked jet
material. For our fiducial model $v'_{\rm s}=0.53$ c. Fender et
al. (1999)\nocite{fgmmpssw99} find that during another outburst of GRS
1915+105 in 1997 $\theta$ was smaller than $8^{\circ}$. This would
then imply an upper limit to $\dot{E}'_{\rm B}$ of $1.2 \cdot 10^{30}$
W. Note that the strength of the magnetic field in the jet after the
passage of the shock is used here. This estimate does not imply that
the unshocked jet material carries a magnetic field of this
strength. Some or all of the magnetic field may be generated in the
shock itself.

Finally, we can derive an lower limit for the bulk kinetic energy
transported by the jet material. We know the number of relativistic
light particles in the jet and so

\begin{eqnarray}
\dot{E}_{\rm kin} & \ge & (\gamma _{\rm j} -1) m_{\rm e} c^2 \dot{N}'_{\rm
o}(R_{\rm o}) e^{-R_{\rm s}/(R_{\rm o} a_4)} \int _{\gamma_{\rm
min}}^{\gamma _{\rm max}} \gamma _{\rm s}^{-p} \, d\gamma _{\rm s}
\nonumber\\
& = &
6.3 \cdot 10^{26} e^{-R_{\rm s}/(R_{\rm o} a_4)} \mbox{W}.
\label{kinetic}
\end{eqnarray}

\noindent This is only a strict lower limit, since we do not know
whether the jets also contain thermal material and/or protons. In the
case that there is one proton for each relativistic electron we find
that the numerical constant in Eq. (\ref{kinetic}) increases to
$1.2 \cdot 10^{30}$. Note that this then is identical to the energy
carried in the form of magnetic fields for $\theta \sim 8^{\circ}$. 

The estimates for the energy transported along the jet in various
forms presented above are lower by about a factor 10 than the
estimates of Fender et al. (1999)\nocite{fgmmpssw99} for the weaker
outburst in 1997. However, it should be noted that their estimates are
based on the assumption that the radio emission is caused by two
`blobs' of relativistic plasma which were ejected by the central
source within about 12 hours. The continuous jet model presented here
requires that the estimated energy supply to the jet is sustained by
the central source for at least 42 days; the length of the first
observing campaign. This means that the total amount of energy
produced by GRS 1915+105 is predicted by our model to be at least an
order of magnitude greater during the March 1994 outburst than it was
in the case of discrete ejections assumed for the September 1997
event.

These estimates illustrate that a continuous jet model cannot decrease
the total amount of energy needed for a given radio outburst but the
rate at which this energy is produced is much lower than in a model
assuming discrete ejection events. This is the case because much, if
not most, of the energy needed to produce the radio emission observed
is `stored' in the material of the continuous jet. This material was
ejected by the central source during comparatively long period well
before the process which led to the formation of the jet shock took
place. Only the acceleration of relativistic particles at the jet
shock then `lights up' the jet and we are able to detect it.

\subsection{Self-absorption}

Since all jet properties are assumed to scale with distance from the
source centre in our conical jet, it is clear that at some early time
in the outburst the jet material was opaque for radio emission because
of synchrotron self-absorption. The absorption coefficient in the rest
frame of the emitting gas is given by (e.g. Longair 1981\nocite{ml81})

\begin{eqnarray}
\chi'_{\nu'} & = & 3.354 \cdot 10^{-9} \left( 3.54
\cdot 10^{18} \right) ^p \nonumber\\
& \times & \kappa ' B'^{(p+2)/2} b(p) \nu'^{(-p-4)/2} \mbox{m}^{-1},
\end{eqnarray}

\noindent where in our notation

\begin{equation}
\kappa ' = \frac{4\dot{N}'_{\rm o}(R_{\rm o}) \left( m_{\rm e} c^2
\right) ^{p-1}}{\pi v'_{\rm s} ( R_{\rm s} \theta)^2} \, e^{-R_{\rm
s}/(R_{\rm o} a_4)},
\end{equation}

\noindent and $b(p)$ is of order unity. We only consider the region
just behind the jet shock where the energy distribution of the
relativistic particles is completely described by a power law of
exponent $p$. For a photon emitted at the centre of the jet the
optical depth in the radial direction is then $\tau = R_{\rm s} \theta
\chi'_{\nu'} / 2$. For our fiducial model we then find that the jet
material becomes transparent at 8.4 GHz roughly 2 hours after the
start of the outburst when the shock has reached a distance of $\sim
2\cdot 10^{12}$ m from the source centre.

Mirabel et al. (1998)\nocite{mdcrmrsg98} find that for the much weaker
`mini-bursts' of GRS 1915+105 the jets become transparent about 30
minutes after the start of the burst. Bearing in mind that the
mini-bursts may be quite different in their properties compared to the
major outburst considered here, our value is therefore in good
agreement with their findings.

\subsection{Infrared emission}

Several groups have reported the detection of infrared emission from
GRS 1915+105 (i.e. Sams et al. 1996\nocite{ses96}, Mirabel et
al. 1996\nocite{mrcsgdcc96}, 1998\nocite{mdcrmrsg98}). In the case of
the mini-bursts simultaneous flux measurements at radio frequencies
and in the K-band are available at times of about 10 to 20 minutes
after the start of the bursts (Mirabel et
al. 1996\nocite{mrcsgdcc96}). Because of the uncertainties in the dust
corrections in the K-band towards GRS 1915+105 it is difficult to
estimate the spectral behaviour from radio to infrared
wavelengths. However, for the mini-bursts flat spectra, $\alpha =0$,
regardless of the exact magnitude of extinction are observed very
early during the bursts (Mirabel et al. 1996\nocite{mrcsgdcc96}). The
slope of the initial energy distribution of the relativistic particles
in combination with the rather low high-energy cut-off of this
distribution we found for our fiducial model is inconsistent with such
flat emission spectra. However, this model does predict an unobscured,
optically thin infrared flux of about 14 mJy in the K-band for a time
about 15 minutes after the start of the burst. This may be enough to
be detected in future observations of large outbursts. The timing
requirements for such an observations are however difficult to meet,
since the predicted infrared flux very quickly becomes undetectable at
only slightly later times.

More puzzling is the detection of a resolved jet component with K-band
flux of at least 1.8 mJy about 0.3" away from the centre of GRS
1915+105 by Sams et al. (1996)\nocite{ses96}. The shock on the
approaching side of our jet model would need 24 days to reach such a
large distance from the source centre. By this time our fiducial model
predicts no synchrotron emission in the K-band at all. We have
estimated whether this infrared emission may be caused by radio
photons which are inverse Compton scattered to such high frequencies
within the jet plasma. However, we find that this cannot explain the
observations since the density of the relativistic particles in the
jet in our model is orders of magnitude too low.

The observation of K-band emission far away from the core of GRS
1915+105 and the flat spectral indices of the mini-bursts suggest that
two different types of outbursts may occur in the jets of this
source. The strong radio bursts like the one of March 1994 are caused
by jet shocks which produce large numbers of relativistic particles
with a steep energy distribution. The weaker mini-bursts involve
shocks which accelerate less particles but produce a flatter energy
distribution which may also extend to higher energies than in the
stronger bursts. The `mini-burst mode' may correspond to a phase of
relative stable jet production with only small variations in the bulk
velocity of the jet material. Such flat spectra extending to
millimeter wavelengths, possibly coupled with the continuous ejection
of a jet, have been observed in Cygnus X-1 (Fender et
al. 2000\nocite{fpdtb00}). The strong radio outbursts then probably
mark phases of more violent changes in the central jet production
mechanisms. Fender (1999)\nocite{rf99} points out that this proposed
behaviour may also be reflected in the X-ray signature of the
accretion disk. In any case, other sources of infrared emission in the
close vicinity of the jets like dust illuminated by the disk and/or
the jet may further complicate the situation (Mirabel et
al. 1996\nocite{mrcsgdcc96}). To test the validity of the proposed
scenario resolved observations of outbursts of GRS 1915+105 and other
galactic jet sources from radio to infrared frequencies would be
necessary.

\section{The end of the jet}
\label{sec:end}

The energy transported by the jets of microquasars is enormous. This
energy will be continuously deposited at the end of the jets and may
lead to significant radiation from this region depending on how it is
dissipated. In the following we investigate the fate of the energy
transported by the jets of microquasars as predicted by our model. The
discussion is based on the work by Leahy (1991)\nocite{jl91}.

\subsection{Momentum balance}
\label{sec:mom}

In order for the jets to expand they have to accelerate the
surrounding ISM and push it aside. The velocity of the contact surface
between the front end of the jet and the ISM, $v_{\rm c}$, is given by
balancing the momentum or `thrust' of the jet material with the ram
pressure of the receding ISM

\begin{eqnarray}
\frac{v_{\rm c}}{v_{\rm j}} & = & \left(1 +\frac{1}{\Gamma _{\rm j}
M_{\rm j}^2} \right) \nonumber \\ & \times & \left[ 1 + \sqrt{
\frac{1}{\eta} \left( 1 + \frac{1}{\Gamma _{\rm j} M_{\rm j}^2}
\right) \left( 1 +\frac{1}{\Gamma _{\rm c} M_{\rm c}^2} \right)
-\frac{1}{\Gamma _{\rm j} M_{\rm j}^2}} \right] ^{-1},
\label{balance}
\end{eqnarray}

\noindent where $\Gamma _{\rm j}$ and $\Gamma _{\rm c}$ are the
adiabatic indices of the jet material and the ISM respectively,
$M_{\rm j}$ is the internal Mach number of the jet flow, $M_{\rm c}$
is the Mach number of the contact surface with respect to the sound
speed in the ISM and $\eta = \rho _{\rm j} / \rho _{\rm c}$. Here
$\rho _{\rm j}$ is the mass density of the jet material while $\rho
_{\rm c}$ is the density of the ISM. This expression is strictly valid
only for non-relativistic jet velocities. However, since the bulk
velocity of the shocked jet material, $v_{\rm j}$, is only mildly
relativistic in our fiducial model, $\gamma _{\rm j} = 1.3$, we take
Eq. (\ref{balance}) to be a good approximation. Note that the
velocity of the jet material in front of the shock is even lower than
$v_{\rm j}$.

For $v_{\rm c} \sim v_{\rm j}$ the jet material does not decelerate
strongly at the end of the jet. This implies that little of the
kinetic energy transported by the jet is dissipated. Even for large
internal Mach numbers it is then unlikely that a strong shock will
develop in the jet flow close to the contact surface. This occurs when
the jet is overdense, i.e. $\eta \gg 1$ and so the jet flow is close
to being ballistic. For underdense jets, $\eta < 1$, the ratio $v_{\rm
c} / v_{\rm j}$ can become considerably smaller than 1. In this case a
strong deceleration of the jet ensues and much of its kinetic energy
is dissipated. For $M_{\rm j} \gg 1$ a strong shock will form and can
act as a site of efficient acceleration of relativistic
particles. Examples for this are the powerful extragalactic radio
sources of type FRII (Fanaroff \& Riley 1974\nocite{fr74}) with their
very bright radio hot spots at the end of their jets. The diffuse
radio lobes enveloping their jets are the remains of the shocked jet
material left behind by the advancing contact surface. In the
transonic regime, $M_{\rm j} \sim 1$, only weak shocks may form at the
jet end and particle acceleration is less efficient. The less powerful
jets of FRI objects fall in this class.

\subsection{Application to our fiducial model}

In the model developed in the previous sections we have assumed the
jets of microquasars to be conical with a constant opening angle. This
implies $\eta = \eta _{\rm o} (R/ R_{\rm o}) ^{-2}$ and, because of
the adiabatic expansion of the jet material, $M_{\rm j} =M_{\rm j}
(R_{\rm o}) (R / R_{\rm o}) ^{\Gamma _{\rm j} -1}$. The bulk velocity
of the jet material is high in our fiducial model and unless the jet
material is very hot ($T_{\rm j} (R_{\rm o})> 10^{12}$ K in the case
of a proton-electron jet) the internal Mach number of the jet flow
will always greatly exceed 1. Since $\eta$ is a strongly decreasing
function of $R$, we expect from Eq. (\ref{balance}) that the ratio
$v_{\rm c}/v_{\rm j}$ will always fall significantly below unity for
large values of $R$. This means that the jets of microquasars should
end eventually in strong shocks which may be detectable in the
radio. In the source XTE J1748-288 a region of bright radio emission
was observed to slow down and brighten at the same time some distance
from the centre of the source (Hjellming et
al. 1999\nocite{hrmshwgp99}). In our model this is interpreted as an
internal shock reaching the end of the jet where the termination shock
further boosts the relativistic particle population which was
pre-accelerated by the internal shock. After passing through the
termination shock the jet material may inflate a radio lobe very
similar to extragalactic FRII objects if $v_{\rm c}/v_{\rm j} \ll 1$
(see also Levinson \& Blandford 1996a,
b\nocite{lb96a}\nocite{lb96b}). It has been suggested that the diffuse
radio emission region W50 around SS433 is the radio lobe inflated by
the jets of this source (Begelman et al. 1980\nocite{bshma80}). Other
radio lobes were detected around 1E 1740.7-2942 (Mirabel et
al. 1992\nocite{mrcpl92}), GRS 1758-258 (Rodr{\'\i}guez et
al. 1992\nocite{rmm92}) and possibly GRO J1655-40 (Hunstead et
al. 1997\nocite{hwc97}), but not in the vicinity of GRS 1915+105
(Rodr{\'\i}guez \& Mirabel 1998\nocite{rm98}). The absence of a radio
lobe in GRS 1915+105 may indicate that the jet in this source is
relatively young and has not yet reached the point at which it becomes
underdense with respect to the ISM. In the following we estimate the
distance out to which the jets in this source may travel without the
formation of a strong termination shock.

A lower limit for $\eta$ can be derived from our fiducial model
assuming that the jets consist only of the relativistic particles
responsible for the synchrotron emission plus the particles needed for
charge neutrality. Thus

\begin{equation}
\eta \ge \dot{N}'_{\rm o} (R_{\rm o}) e^{-R_{\rm s} / (R_{\rm o} a_4)}
\frac{m_{\rm j}}{\pi \left( \theta /2 R_{\rm s} \right) ^2 v'_{\rm s}
\rho _{\rm c}} \int_{\gamma _{\rm min}}^{\gamma _{\rm max}} \gamma
_{\rm s} ^{-p} \, d\gamma _{\rm s},
\label{lower}
\end{equation}

\noindent where $m_{\rm j}$ is the mass of the average particle in the
jet. An upper limit for $\eta$ can be derived from the assumption that
the rate at which mass is ejected along the jet can not exceed the
mass accretion rate within the disk powering the jet. Fits to the
X-ray spectrum of GRS 1915+105 suggest an accretion rate of order
$\dot{m} \sim 10^{15}$ kg s$^{-1}$ (Belloni et
al. 1997\nocite{bmkkp97}). We then find

\begin{equation}
\eta \le \frac{\dot{m}}{\pi \left( \theta / 2 R_{\rm s} \right) ^2
v_{\rm s}' \rho _{\rm c}}.
\label{upper}
\end{equation}

\noindent Note that this upper limit does not depend on the nature of
the jet material. Using Eqs. (\ref{lower}) and (\ref{upper}) and
assuming $\rho _{\rm c} \sim m_{\rm p} 10^6$ kg m$^{-3}$,
corresponding to a particle density of 1 cm$^{-3}$, we find $56 \le
\eta _{\rm o} \le 1300$ for a proton-electron jet and $0.03 \le \eta
_{\rm o} \le 1300$ for a pair plasma jet. The lower limit for the pair
plasma jet assumes that the pairs are cold. Because of pair
annihilation it is unlikely that the material of a pair plasma jet is
cold (e.g. Gliozzi et al. 1999\nocite{gbg99}) and so this lower limit
is used here for illustrative purposes only. Relativistic thermal
motion of the pairs would raise this lower limit.

\begin{figure}
\centerline{
\epsfig{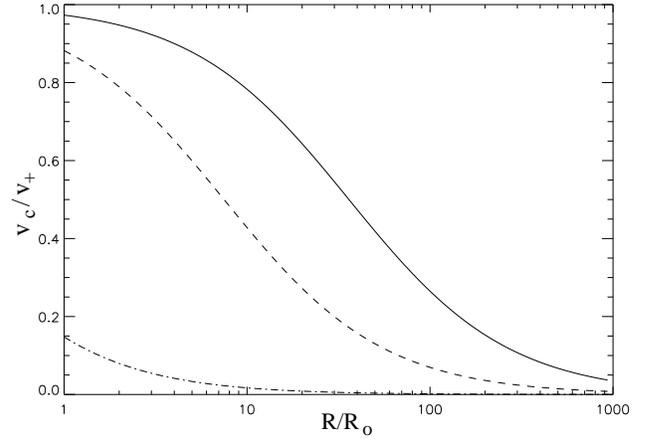}}
\caption{The advance speed of the termination shock at the end of a
hypersonic, conical jet in units of the bulk velocity of the jet
material. Solid line: The case of maximum mass transport rate. Dashed
line: Minimum mass transport rate of a proton-electron jet. Dot-dashed
line: Minimum mass transport rate for a pair plasma jet.}
\label{fig:balance} 
\end{figure}

For the reasonable assumptions $M_{\rm c} \gg 1$ and $M_{\rm j} \gg 1$
Fig. \ref{fig:balance} shows the ratio $v_{\rm c} / v_{\rm j}$ as
calculated from Eq. (\ref{balance}). We see that even if the mass
transport rate of the jet is equal to the mass accretion rate a
termination shock should form about $50 R_{\rm o} \sim 5\cdot 10^{15}$
m away from the core of the source. This distance is reached by the
jet material traveling at $v_{\rm j}=0.61$ c in less than two years.

GRS 1915+105 was discovered as a bright X-ray source on the 15th of
August 1992 (Castro-Tirado et al. 1992\nocite{cbl92}). Given the
availability of X-ray monitoring satellites before 1992 it is not
likely that this source was very active before this date. The
subsequent radio monitoring with the Green Bank Interferometer
(e.g. Foster et al. 1996\nocite{fwthzpg96}) shows that after its
discovery GRS 1915+105 produced radio outbursts every few months. In
the frame of the internal shock model described here this implies that
jet production must have been reasonably steady since 1992. Assuming
that the bulk velocity of the jet material did not vary strongly, the
end of the jet must have reached a distance of roughly $3\cdot
10^{16}$ m from the core by the end of 1997. This is the time of the
radio observation of the large scale surroundings of GRS 1915+105 by
Rodr{\'\i}guez \& Mirabel (1998)\nocite{rm98} who did not find any
evidence for a termination shock of the jet or radio lobes (but see
Levinson \& Blandford 1996a\nocite{lb96a}).

The estimation of the position of the termination shock depends
crucially on the overdensity of the jet material with respect to the
ISM. It is possible that the gas density in the vicinity of GRS
1915+105 is lower than assumed here. However, given its location in
the galactic plane this is rather unlikely. A further possibility is
that the jets of microquasars are not conical for their entire
length. The jets of extragalactic FRII objects are believed to pass
through a very oblique reconfinement shock which brings them into
pressure equilibrium with their environment (e.g. Falle
1991\nocite{sf91}). These shocks are not very efficient in
accelerating relativistic particles and so are often
undetectable. This scenario is also confirmed for FRII sources by
numerical simulations of their jets (e.g. Komissarov \& Falle
1998\nocite{kf98}). The same process may recollimate the jets of
microquasars as well. In this case they may stay overdense with
respect to the ISM much longer and this would enable them to travel
out to much larger distances before terminating in a strong shock. In
this respect it is interesting to note that Rodr{\'\i}guez \& Mirabel
(1998)\nocite{rm98} found a compact non-thermal emission region
located 16.3' away from GRS 1915+105. The feature is elongated and its
major axis is aligned with one of the jets. If this feature is caused
by the jet pointing in its direction then it must have been ejected by
the core roughly 280 years ago. This may be the time scale on which
GRS 1915+105 becomes active and produces jets.

\section{Observational tests of the model}
\label{sec:test}

\subsection{Emission lines}

In our model the jet flow initially consists of non-relativistic hot
plasma. Due to adiabatic expansion losses of thermal energy during the
propagation of this material along the conical jet with given opening
angle, the temperature of the plasma decreases. Internal shocks as
envisioned above lead to local heating and acceleration of
relativistic particles but do not change this general picture. In
reality the situation is very similar to that in the well-known source
SS433. In SS433 we observe bright optical recombination Balmer,
Paschen and Brackett lines of hydrogen which are blue-shifted in the
approaching jet and red-shifted in the receding jet. The velocity of
the jet flow in SS433 is equal to 0.26 c. Due to the precession of the
jet the observed red and blue shifts are strong functions of time
(Margon 1984\nocite{bm84}). In our case the inclination angle of the
jets of GRS 1915+105 is known from the observations by Mirabel \&
Rodr{\'\i}guez (1994)\nocite{mr94}. This angle, $\theta =68^{\circ}$,
and the bulk velocity of 0.6 c of the jet material found in our
fiducial model permits us to estimate the red and blue line shifts of
the emitting jet material:

\begin{eqnarray}
\frac{\lambda}{\lambda '} = \gamma _j (1 \pm \beta _j \cos \theta) & =
& \left\{
\begin{array}{ll}
0.97 & ; {\rm approaching}\\
1.53 & ; {\rm receding}
\end{array}
\right.
\end{eqnarray}

\noindent Note that any line emission coming from the jet approaching
the observer is hardly shifted in wavelength at all. Assuming that the
bulk velocity of the jet material in the jets of GRO J1655-40 is also
close to 0.6 c, we find for this source that the emission lines are
redshifted for the approaching jet ($\lambda / \lambda ' \sim 1.18$)
as well as for the receding jet ($\lambda / \lambda ' \sim
1.32$). This is caused by the large viewing angle, $\theta
=85^{\circ}$, of the jets in this object (Hjellming \& Rupen
1995\nocite{hr95}). In both cases the very large inclination angles
result in a strong predicted asymmetry in the line shifts for the two
jets. Measuring these shifts will permit us to estimate both the
velocity of the jet bulk flow and the viewing angle
$\theta$. Furthermore, any jet precession as in the case of SS433 could
be detected.

Measuring the predicted line shifts is complicated by the low density
of the material in the jet flow of SS433 and GRS 1915+105 which
prevents the production of bright recombination lines. However, we
know that in the case of SS433 there is a strong thermal instability
in the flow which leads to the formation of small, dense cloudlets
(Panferov \& Fabrika 1997\nocite{pf97}). This increases the
recombination rate and effective emission measure of the plasma in the
flow. If there is a similar instability in the jets of GRS 1915+105
and GRO J1655-40 we have a good chance to observe recombination
lines from both of these sources. Another problem is the strong
obscuration of GRS 1915+105 by interstellar dust. Therefore, it is
only possible to look for recombination lines of hydrogen in the
K-band. In the case of GRO J1655-40 obscuration is low and there is a
chance to detect Lyman and Balmer lines. Unfortunately, the mechanical
power of the jet in GRO J1655-40 is smaller than in GRS 1915+105 or
SS433. This will lead to a smaller density and emission measure of the
jet material and therefore also a smaller intensity of the lines. It
is important to bear in mind that in SS433 the emission lines are
extraordinarily bright and but modern observational techniques permit
us to look for blue and red-shifted lines which are weaker by many
orders of magnitude.

In SS433 ASCA discovered red and blue-shifted X-ray K-lines of iron
with a rest energy of roughly 6.7 keV and similar lines of hydrogen-
and helium-like sulphur and argon (Kotani et
al. 1997\nocite{kkmb97}). In our case the cooling jet flow with an
initially very high temperature must lead to the emission in similar
lines of recombining high-Z ions. Again, the mechanical energy of the
flows in GRS 1915+105 and GRO J1655-40 is smaller than in SS433 and,
therefore, the lines should be weaker in these objects. However, the
new X-ray spacecraft, XMM, CHANDRA, ASTRO-E, Constellation-X and XEUS,
may be able to detect such emission in red and blue-shifted X-ray
lines. Note in this respect the detection of shifted iron lines in GRO
J1655-40 reported by Ba\l uci{\'n}ska-Church \& Church
(2000)\nocite{bc00} with RXTE which the authors attribute to the
accretion disk but may very well originate in the continuous jets of
this source.

All predictions for the production of line emission in the jets are
based on the assumption that the jet flow consists of matter with a
high but non-relativistic temperature moving as a whole with
relativistic bulk velocities. There are two other obvious
possibilities: (i) The jet matter consists of a pair plasma and (ii)
the jets consist only of ultra-relativistic plasma with no cold
electrons present. In case (i) we have to consider the possibility of
a bubble around an X-ray source filled with a huge amount of
positrons. If these positrons become non-relativistic due to adiabatic
or other energy losses inside the jets and they cool down to
sufficiently low temperatures, we may observe a blue and red-shifted
recombination line of positronium in the optical and UV
wavebands. This line has a wavelength twice that of the Ly-$\alpha$
line of hydrogen. Much more important in this case, annihilation lines
could be observed again red and blue-shifted relative to the rest
energy of 511 keV. The strong red-shift but weak blue-shift of this
line predicted by our model leaves a unique signature which will be
observable with INTEGRAL. A luminosity only a few times smaller than
the mechanical power of the jets will be emitted in the
electron-positron annihilation line in this case. This large
luminosity should make the annihilation lines observable despite the
unfavorable angle of the jets to our line of sight. In the case of
only ultra-relativistic plasma in the jets, case (ii), no
recombination or annihilation lines should be observable.

\subsection{Radio continuum}

The internal shock models of GRBs (Rees \& Meszaros 1994\nocite{rm94})
attribute the formation of the shock traveling along the jet to the
collision of shells of jet material with different bulk velocities. In
the non-relativistic limit the velocity of the resulting shock is
governed by the same momentum balance, Eq. (\ref{balance}), as the
velocity of the termination surface of the jet. {\bf All quantities in
that equation with subscript `c' now refer to the slower jet material
in front of the jet shock while those with subscript `j' denote
properties of the faster jet material driving the shock. The density
ratio $\eta$ is now simply given by the densities of the faster jet
material driving the shock, $\rho _2$, and that of the slower gas in
front of the shock, $\rho _1$. We already pointed out in
Sect. \ref{sec:mom} that the jet shock is likely to be strong and so
both Mach numbers in Eq. \ref{balance} are significantly greater than
unity. Therefore $v_{\rm s} \sim v_{\rm j} \left( 1 + 1 / \sqrt{\eta}
\right) ^{-1}$. Since in our model all material is assumed to be part
of the conical jet structure, we find $\rho _1 \propto \rho _2 \propto
R^{-2}$ and therefore $\eta = \rho _1 /\rho _2 ={\rm const.}$. This
implies that within the limitations of the model presented here the
velocity of the shock is constant as well which is confirmed by the
observations (e.g. Mirabel \& Rodr{\'\i}guez 1999\nocite{mr99} and
references therein).}

Once the energy of the shell collision is spent, the shock emission
fades rapidly. It is therefore possible that we can observe the shock
reaching the end of the jet only in special cases (XTE J1748-288,
Hjellming et al. 1999\nocite{hrmshwgp99}; see above). We would then
expect that the superluminal component should brighten, as well as
decelerate rather abruptly. In the plasmon model the observed constant
superluminal motion is taken to indicate a large mass and consequently
large kinetic energy of the plasmon. If a plasmon is observed to slow
down because of the growing mass of ISM it sweeps up, then this
deceleration should be rather gradual unless the plasmon encounters a
local overdensity in the ISM. The observed deceleration of the
superluminal component in XTE J1748-288 occurred rather rapidly at a
distance of about 1" from the core after a phase of expansion with
practically constant velocity. Furthermore, the emission region is
still detected in recent observations; 15 months after the start of
the burst (Rupen, private communication). During this time it appears
to have advanced only slowly at a velocity of about 0.01" per month or
roughly 5000 km s$^{-1}$. This slow motion and persistent radio
emission may be interpreted as arising from the shock at the end of a
continuous jet (see the previous section).

\begin{figure}
\centerline{
\epsfig{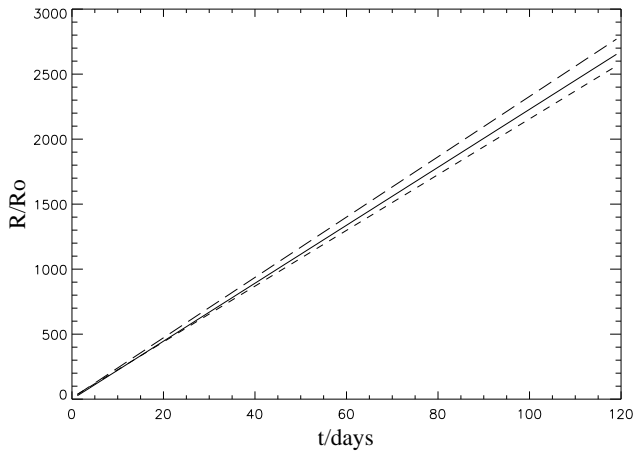}}
\caption{Position of the radio emission peak in the fiducial model as
a function of time after the start of the outburst. Solid line:
Velocity of the jet shock derived from the 8.4 GHz observations,
dashed line: Model prediction at 1.4 GHz and long dashed line: Model
prediction at 8.4 GHz. Only the approaching jet component is plotted.}
\label{fig:movpeak}
\end{figure}

Some interesting predictions can be made from the model for future
radio observations in the case that these can resolve the approaching
and receding jet components along the jet axis. Because of the way in
which the rate of acceleration of relativistic particles in the jet by
the shock varies with time, the peak of the radio emission is not
coincident with the position of the shock. This off-set depends on the
observing frequency in the sense that the lower this frequency the
more the emission peak lags behind the leading shock. This is
illustrated in Fig. \ref{fig:movpeak} where we plot the distance of
the emission peak on the approaching jet side as a function of time
for two different frequencies. Note also that this effect predicts
that we should measure slightly different advance velocities of the
emission peaks at different frequencies. This will not be observed in
the case of discrete plasmon ejections.

\begin{figure}
\centerline{
\epsfig{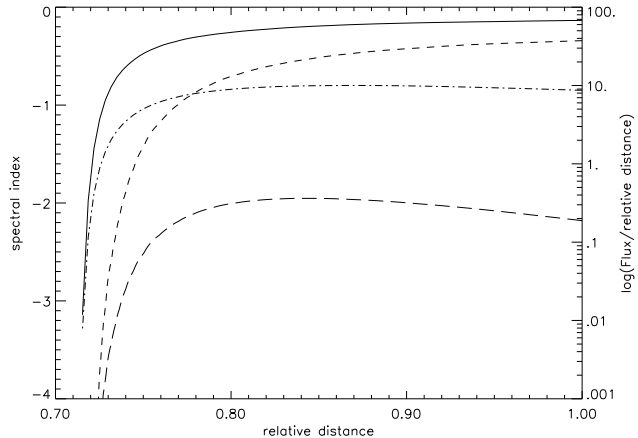}}
\caption{Predicted variation of the radio spectral index between 1.4
GHz and 8.4 GHz and flux along the jet axis. The length scale on the
x-axis is normalised to the distance of the jet shock from the source
centre at the respective observing time. Solid line: At the time of
the first VLA observations, i.e. 4.8 days after the start of the
outburst. Short-dashed line: At a time of 40 days after the start of
the outburst. Dot-dashed line: Flux per relative distance in arbitrary
units at 8.4 GHz at the time of the first VLA observations,
corresponding to solid line. Long-dashed line: Flux per relative
distance 40 days after the start of the outburst, corresponding to
short-dashed line.}
\label{fig:spec}
\end{figure}

The steepening of the radio spectrum of the jets in microquasars in
this model is explained by the superposition of the contribution to
the total emission from various regions within the jet. In resolved
radio maps of the jet components this should be visible because the
model predicts the radio spectral index to change along the jet
axis. This behaviour is shown in Fig. \ref{fig:spec} for the
approaching jet. {\bf Fig. \ref{fig:spec} also shows the distribution
of the flux along the jet axis. The relatively uniform distribution is
caused by the decrease of the magnetic field strength further out
along the jet counteracting the injection of newly accelerated
particles by the shock. The jet region over which the spectrum
steepens is small and the emission originating in this region also
weakens considerably in the direction away from the jet shock. This
may make a detection of the spectral steepening along the jet
difficult. However, the apparent shortening of the emission region
along the jet at higher observing frequencies may be detectable.} The
decrease in the strength of the magnetic field combined with the high
energy cut-off of the energy spectrum of the relativistic particles
leads to an overall steepening with time of the radio spectrum along
the jet axis. This is also shown in Fig. \ref{fig:spec} and should be
observable if the jet components can be resolved at more than one
frequency.

Note also that the length of the region along the jet axis which is
emitting radiation at a given frequency increases with time. Although
the fraction of the distance of the shock from the source centre
subtended by the emitting region shrinks for later times (see
Fig. \ref{fig:spec}), the absolute extent of this region will
grow. This may also be detectable in future radio observations of
sufficient surface brightness sensitivity.

Another prediction of the model is that the lightcurves of radio
outbursts in microquasars at a given observing frequency should have a
fairly constant slope for a few tens of days. After that they steepen
rapidly once the critical frequency of the most energetic relativistic
particles moves below the observing frequency. This steepening occurs
earlier at higher frequencies. At the same time that the steepening of
the lightcurves occurs, the flux ratio of the approaching and receding
jet components should decrease. This effect is seen in Fig.
\ref{fig:res8.4}. The jet components of the second outburst of April
21 observed during the second observing campaign with the VLA
(Rodr{\'\i}guez \& Mirabel 1999\nocite{rm99}) show a much steeper
lightcurve than those of the first outburst and, at the same time, a
smaller flux ratio. Since these components were observed later in
their evolution than those of the first outburst, this is in agreement
with the predictions of the model.

\section{Conclusions}

We developed a continuous jet model for the radio outbursts of
galactic microquasars. The model naturally explains the observed
rather flat decaying lightcurves of these bursts as the signature of
synchrotron radiation of relativistic particles accelerated by
internal shocks in the conical jets. The comparatively long duration
of the bursts implies that this model is a `time-resolved' version of
the internal shock model proposed for GRB (Rees \& Meszaros
1994\nocite{rm94}), though the synchrotron emission is produced at
much lower frequencies. The gradual steepening of the radio spectrum
is explained by a superposition of the radiation of different
populations of relativistic particles with different ages. This
spectrum of ages results from the shock traveling along the jet with
older populations of accelerated particles left behind.

We find that only a roughly constant rate of acceleration of
relativistic particles followed by an exponential decay can explain
the observed light curves for the strong outburst of GRS 1915+105 in
1994. We interpret this behaviour as the signature of two colliding
shells of jet material, as in the internal shock model for GRB. A
consequence of this is the continued steepening of the lightcurves of
a given outburst coupled with a decreasing flux ratio of the emission
observed from the approaching to that from the receding jet side.

The energy requirements of the continuous jet model for producing
radio outbursts are similar to those of the plasmon model. However,
much of the energy underlying the outbursts may be stored in the
continuous jet while the passage of the internal shock only `lights
up' the jet. This implies that the rate at which the energy of the
outburst is supplied by the central engine to the jet is much lower
than in the plasmon model. 

The occurrence of mini-bursts in microquasars with flat spectra up to
infrared frequencies (Mirabel et al. 1998\nocite{mdcrmrsg98}) and the
observation of K-band emission in the jet a considerable distance away
from the core (Sams et al. 1996\nocite{ses96}) suggest different modes
of jet production: (i) A stable `mini-burst' mode with relative little
variation in the bulk jet speed and therefore also only weak internal
shocks. (ii) A more variable outburst mode with strong variations in
the jet speed and strong internal shocks (see also Fender
1999\nocite{rf99}). The weaker flavour internal shocks seem to produce
flatter relativistic particle spectra extending to higher energy
compared to the strong shocks. However, the total number of
accelerated particles must be much larger in the strongly variable
phase.

We show that the properties of the continuous jets of microquasars
should lead to strong shocks at their ends where they are in contact
with the surrounding ISM. This is consistent with the recent
observations of the decelerating radio emission region of XTE
J1748-288 (Hjellming et al. 1999\nocite{hrmshwgp99}) and its
persistence for 15 months after the start of the original outburst
(Rupen, private communication). The shocked jet material may
subsequently inflate a low density cavity around the jets similar to
the radio lobes in extragalactic jet sources of type FRII. This is
observed in SS433 (Dubner et al. 1998\nocite{dhgm98}) and may be in a
few other microquasars. The absence of such shocks and radio lobes in
GRS 1915+105 may indicate that the jets in this source are young
and/or that they recollimate because of the pressure of their
environment. If this is the case then the detection of a non-thermal
emission region in the more extended environment of GRS 1915+105
(Rodr{\'\i}guez \& Mirabel 1998\nocite{rm98}) may imply a recurrence
time of the jet activity scale of $\sim 280$ years.

Many of the predictions of this model for microquasars can be tested
observationally. However, to clearly distinguish between this model of
continuous jets and the plasmon model it would be necessary to
spatially resolve the superluminal emission regions during outbursts,
preferentially at more than one radio frequency. Additional support
for the scenario of continuous jets may come from further high
resolution observations of the cores of microquasars during
quiescence. These should show at least some spatial extension of the
radio emission along the jet axis as observed in Cygnus X-1 (Fender et
al. 2000\nocite{fpdtb00}). These observations can potentially provide
us with valuable information on the properties of the jets which
otherwise we can only study during strong outbursts when strong shocks
pass through them.

\section*{Acknowledgments}

The authors would like to thank G. Ghisellini and the referee,
L.F. Rodr{\'\i}guez, for valuable discussions which improved the
paper. This work was partly supported by EC grant ERB-CHRX-CT93-0329
within the research network `Accretion onto compact objects and
proto-stars'


\begin{thebibliography}{50}

\bibitem[\protect\astroncite{Atoyan \& Aharonian}{1999}]{aa99}
Atoyan A.M., Aharonian F.A., 1999, MNRAS 302, 253

\bibitem[\protect\astroncite{{Ba\l uci\'nska-Church} \&
Church}{2000}]{bc00} {Ba\l uci\'nska-Church} M., Church M.J., 2000,
MNRAS: accepted, astro-ph/9912389.

\bibitem[\protect\astroncite{Begelman et~al.}{1980}]{bshma80}
Begelman M.C., Sarazin C.L., Hatchett S.P., McKee C.F., Arons J., 1980, ApJ
  238, 722

\bibitem[\protect\astroncite{Belloni et~al.}{1997}]{bmkkp97}
Belloni T., Mendez M., King A.R., van~der Klis M., Paradijs J.V., 1997, ApJ
  488, L109

\bibitem[\protect\astroncite{Blandford \& Rees}{1974}]{br74}
Blandford R.D., Rees M.J., 1974, MNRAS 169, 395

\bibitem[\protect\astroncite{Castro-Tirado et~al.}{1992}]{cbl92}
Castro-Tirado A., Brandt S., Lund N., 1992, IAU Circ 5590

\bibitem[\protect\astroncite{Drury}{1983}]{ld83}
Drury L.{\protect{O'C}}., 1983, Rep. Prog. Phys. 46, 973

\bibitem[\protect\astroncite{Dubner et~al.}{1998}]{dhgm98}
Dubner G.M., Holdaway M., Goss W.M., Mirabel I.F., 1998, AJ 116, 1842

\bibitem[\protect\astroncite{Falle}{1991}]{sf91}
Falle S.A.E.G., 1991, MNRAS 250, 581

\bibitem[\protect\astroncite{Fanaroff \& Riley}{1974}]{fr74}
Fanaroff B.L., Riley J.M., 1974, MNRAS 167, 31

\bibitem[\protect\astroncite{Fender}{1999}]{rf99}
Fender R.P., 1999, in L. Kaper, E.~P.~J. van~den Heuvel, and P.~A. Woudt
  (eds.), Black holes in binaries and galactic nuclei. Springer, p.~in press

\bibitem[\protect\astroncite{Fender et~al.}{1999}]{fgmmpssw99}
Fender R.P., Garrington S.T., McKay D.J. et~al., 1999, MNRAS 304, 865

\bibitem[\protect\astroncite{Fender et~al.}{2000}]{fpdtb00} Fender
R.P., Pooley G.G., Durouchoux P., Tilanus R.P.J., Brocksopp C., 2000,
MNRAS: accepted, astro-ph/9910184.

\bibitem[\protect\astroncite{Foster et~al.}{1996}]{fwthzpg96}
Foster R.S., Waltman E.B., Tavani M. et~al., 1996, ApJ 467, L81

\bibitem[\protect\astroncite{Ghisellini}{1999}]{gg99}
Ghisellini G., 1999, Astron. Nach., submitted, astro-ph/9906145

\bibitem[\protect\astroncite{Ghisellini et~al.}{1985}]{gmt85}
Ghisellini G., Maraschi L., Treves A., 1985, A\&A 146, 204

\bibitem[\protect\astroncite{Gliozzi et~al.}{1999}]{gbg99}
Gliozzi M., Bodo G., Ghisellini G., 1999, MNRAS 303, L37

\bibitem[\protect\astroncite{Heavens \& Meisenheimer}{1987}]{hm87}
Heavens A.F., Meisenheimer K., 1987, MNRAS 225, 335

\bibitem[\protect\astroncite{Hjellming \& Johnston}{1988}]{hj88}
Hjellming R.M., Johnston K.J., 1988, ApJ 328, 600

\bibitem[\protect\astroncite{Hjellming \& Rupen}{1995}]{hr95}
Hjellming R.M., Rupen M.P., 1995, Nat 375, 464

\bibitem[\protect\astroncite{Hjellming et~al.}{1999}]{hrmshwgp99}
Hjellming R.M., Rupen M.P., Mioduszewski A.J. et~al., 1999, American
  Astronomical Society Meeting 193, 103.08

\bibitem[\protect\astroncite{Hunstead et~al.}{1997}]{hwc97}
Hunstead R.W., Wu K., Campbell-Wilson D., 1997, in D.~T. Wickramasinghe, L.
  Ferrario, and G.~V. Bicknell (eds.), Accretion phenomena and related
  outflows. ASP Conference Series, p.~63

\bibitem[\protect\astroncite{Kaiser et~al.}{1997}]{kda97a}
Kaiser C.R., Dennett-Thorpe J., Alexander P., 1997, MNRAS 292, 723

\bibitem[\protect\astroncite{Komissarov \& Falle}{1998}]{kf98}
Komissarov S.S., Falle S.A.E.G., 1998, MNRAS 297, 1087

\bibitem[\protect\astroncite{Kotani et~al.}{1997}]{kkmb97}
Kotani T., Kawai N., Matsuoka M., Brinkmann W., 1997, in D. Wickramasinghe, L.
  Ferrario, and G. Bicknell (eds.), Accretion phenomena and related outflows.
  ASP Conference series, p.~270

\bibitem[\protect\astroncite{Leahy}{1991}]{jl91}
Leahy J.P., 1991, in P.~A. Hughes (ed.), Beams and jets in astrophysics.
  Cambridge University Press, p.~100

\bibitem[\protect\astroncite{Levinson \& Blandford}{1996{a}}]{lb96b}
Levinson A., Blandford R., 1996{a}, A\&A Supp. 120, 129

\bibitem[\protect\astroncite{Levinson \& Blandford}{1996{b}}]{lb96a}
Levinson A., Blandford R., 1996{b}, ApJ 456, L29

\bibitem[\protect\astroncite{Longair}{1981}]{ml81}
Longair M.S., 1981, {\em High energy astrophysics\/}, Cambridge University
  Press

\bibitem[\protect\astroncite{Margon}{1984}]{bm84}
Margon B., 1984, ARA\&A 22, 507

\bibitem[\protect\astroncite{Marscher \& Gear}{1985}]{mg85}
Marscher A.P., Gear W.K., 1985, ApJ 298, 114

\bibitem[\protect\astroncite{Mirabel \& Rodr{\'\i}guez}{1994}]{mr94}
Mirabel I.F., Rodr{\'\i}guez L.F., 1994, Nat 371, 46

\bibitem[\protect\astroncite{Mirabel \& Rodr{\'\i}guez}{1998}]{mr98}
Mirabel I.F., Rodr{\'\i}guez L.F., 1998, Nat 392, 673

\bibitem[\protect\astroncite{Mirabel \& Rodr{\'\i}guez}{1999}]{mr99}
Mirabel I.F., Rodr{\'\i}guez L.F., 1999, ARA\&A 37, 409

\bibitem[\protect\astroncite{Mirabel et~al.}{1992}]{mrcpl92}
Mirabel I.F., Rodr{\'\i}guez L.F., Cordier B., Paul J., Lebrun F., 1992, Nat
  358, 215

\bibitem[\protect\astroncite{Mirabel et~al.}{1996}]{mrcsgdcc96}
Mirabel I.F., Rodr{\'\i}guez L.F., Chaty S. et~al., 1996, ApJ 472, L111

\bibitem[\protect\astroncite{Mirabel et~al.}{1998}]{mdcrmrsg98}
Mirabel I.F., Dhawan V., Chaty S. et~al., 1998, A\&A 330, L9

\bibitem[\protect\astroncite{Panferov \& Fabrika}{1997}]{pf97}
Panferov A.A., Fabrika S.N., 1997, Astronomy Reports 41, 506

\bibitem[\protect\astroncite{Press et~al.}{1992}]{ptvf92}
Press W.H., Teukolsky S.A., Vetterling W.T., Flannery B.P., 1992, {\em
  Numerical Recipes. Second edition.\/}, Cambridge University Press, Cambridge,
  UK.

\bibitem[\protect\astroncite{Rees}{1978}]{mr78}
Rees M.J., 1978, MNRAS 184, P61

\bibitem[\protect\astroncite{Rees \& Meszaros}{1994}]{rm94}
Rees M.J., Meszaros P., 1994, ApJ 430, L93

\bibitem[\protect\astroncite{Reynolds \& Begelman}{1997}]{rb97}
Reynolds C.S., Begelman M.C., 1997, ApJ 487, L135

\bibitem[\protect\astroncite{Rodr{\'\i}guez \& Mirabel}{1998}]{rm98}
Rodr{\'\i}guez L.F., Mirabel I.F., 1998, A\&A 340, L47

\bibitem[\protect\astroncite{Rodr{\'\i}guez \& Mirabel}{1999}]{rm99}
Rodr{\'\i}guez L.F., Mirabel I.F., 1999, ApJ 511, 398

\bibitem[\protect\astroncite{Rodr{\'\i}guez et~al.}{1992}]{rmm92}
Rodr{\'\i}guez L.F., Mirabel I.F., Mart{\'\i} J., 1992, ApJ 401, L15

\bibitem[\protect\astroncite{Rodr{\'\i}guez et~al.}{1995}]{rgmgv95}
Rodr{\'\i}guez L.F., Gerard E., Mirabel I.F., G{\'o}mez Y., Vel{\'a}zquez A.,
  1995, ApJ 101, 173

\bibitem[\protect\astroncite{Rybicki \& Lightman}{1979}]{rl79}
Rybicki G.B., Lightman A., 1979, {\em Radiative Processes in Astrophysics\/},
  Wiley, New York.

\bibitem[\protect\astroncite{Sams et~al.}{1996}]{ses96}
Sams B.J., Eckart A., Sunyaev R., 1996, Nat 382, 47

\bibitem[\protect\astroncite{Shu}{1991}]{fs91}
Shu F.H., 1991, {\em The physics of astrophysics. Vol.1: Radiation\/},
  University Science Books, Mill Valley

\bibitem[\protect\astroncite{van Paradijs}{1995}]{jv95}
van Paradijs J., 1995, in W.~H.~G. Lewin, J. van Paradijs, and E.~P.~J. van~den
  Heuvel (eds.), X-ray binaries. Cambridge University Press, p.~536

\end{thebibliography}
\end{document}